\documentclass[aip,apl,reprint]{revtex4-1}
\usepackage{gensymb}
\usepackage{graphicx}
\usepackage{textcomp}
\usepackage{verbatim}
\usepackage{setspace}
\usepackage{textcomp}
\usepackage{bm}
\usepackage{amsmath}
\usepackage[colorlinks=true,urlcolor=blue,linkcolor=blue,citecolor=blue,a4paper]{hyperref}
\graphicspath{{./images/}}
\begin{document}

\title{High-temperature-grown buffer layer boosts electron mobility in epitaxial La-doped BaSnO$_3$/SrZrO$_3$ heterostructures}

\author{Arnaud P. Nono Tchiomo}
\affiliation{Max Planck Institute for Solid State Research, Heisenbergstr.\ 1, 70569 Stuttgart, Germany}
\affiliation{Department of Physics, University of Johannesburg, P.O. Box 524 Auckland Park 2006, Johannesburg, South Africa}
\author{Wolfgang Braun}
\affiliation{Max Planck Institute for Solid State Research, Heisenbergstr.\ 1, 70569 Stuttgart, Germany}
\author{Bryan P. Doyle}
\affiliation{Department of Physics, University of Johannesburg, P.O. Box 524 Auckland Park 2006, Johannesburg, South Africa}
\author{Wilfried Sigle}
\affiliation{Max Planck Institute for Solid State Research, Heisenbergstr.\ 1, 70569 Stuttgart, Germany}
\author{Peter van Aken}
\affiliation{Max Planck Institute for Solid State Research, Heisenbergstr.\ 1, 70569 Stuttgart, Germany}
\author{Jochen Mannhart}
\affiliation{Max Planck Institute for Solid State Research, Heisenbergstr.\ 1, 70569 Stuttgart, Germany}
\author{Prosper Ngabonziza}
\email[Corresponding author: ]{ p.ngabonziza@fkf.mpg.de}
\affiliation{Max Planck Institute for Solid State Research, Heisenbergstr.\ 1, 70569 Stuttgart, Germany}
\affiliation{Department of Physics, University of Johannesburg, P.O. Box 524 Auckland Park 2006, Johannesburg, South Africa}
\email[corresponding author, ]{p.ngabonziza@fkf.mpg.de}
\date{\today}

\begin{abstract}
By inserting a SrZrO$_3$  buffer layer between the film and the substrate, we demonstrate a significant reduction of the threading dislocation density with an associated improvement of the electron mobility in La:BaSnO$_3$ films.  A room temperature mobility of 140 cm$^2$ V$^{-1}\text{s}^{-1}$ is achieved for 25-nm-thick films without any post-growth treatment. The density of threading dislocations is only $4.9\times 10^{9}$ cm$^{-2}$ for buffered films prepared on (110) TbScO$_3$ substrates by pulsed laser deposition. 
\end{abstract}
\maketitle
Transparent conducting oxides (TCOs) have attracted attention due to their unique properties and applications in electronic devices such as transparent displays and transistors.\cite{HOhta2003,WJLee2017} Recently, the transparent perovskite La:BaSnO$_3$ has gained interest as a novel TCO due to its high mobility at room temperature (RT).\cite{HJKim2012-1,HJKim2012-2} Single crystals of La:BaSnO$_3$ have been reported to have RT mobilities as high as 320 cm$^2$ V$^{-1}\text{s}^{-1}$ (mobile carrier concentration $n=8\times 10^{19}\, $cm$^{-3}$).\cite{HJKim2012-2} At doping concentrations above $10^{19}\, $cm$^{-3}$, La:BaSnO$_3$ has the second highest RT electron mobility among TCOs and other oxide single crystals~\cite{HJKim2012-2,XLuo2012,HMIJaim2017}, exceeded only by CdO.~\cite{ESachet2015}  The high RT mobility in La:BaSnO$_3$ has been attributed to the small electron-phonon interaction and small electron effective mass ($m_e^*=0.19\; m_0$) arising from the large dispersion of the conduction band comprised of Sn 5s orbitals.\cite{APrakash2017,KKrishnaswamy2017,CANiedermeier2017} However, to fully explore the potential of La:BaSnO$_3$, thin films with high mobilities are also required.

Recently, several efforts to achieve high carrier mobility in La:BaSnO$_3$ films have been reported.~\cite{JYue2018,KFujiwara2016,CPark2014,UKim2015,JShin2016,SArezoomandan2018,HPaik2017,ZWang2019}  Nevertheless, as compared to bulk single crystals, even the best La:BaSnO$_3$ epitaxial films show a reduced mobility ($\leq 183$ cm$^2$ V$^{-1}\text{s}^{-1}$),\cite{HPaik2017} which specifically has been attributed to scattering from charge defects, such as threading dislocations (TDs)\cite{APrakash2017,HPaik2017,SRaghavana2016,HMun2013} and, more generally, to small carrier relaxation times.\cite{APrakash2017,KKrishnaswamy2017} TDs form for a large lattice mismatch between film and substrate, and extend perpendicularly through the films.  The obvious solution to  reduce TDs in La:BaSnO$_3$ films is to use lattice-matched substrates, but unfortunately none exists.  The substrate with the closest lattice match that is commercially available is PrScO$_3$, mismatched by $- 2.3$\%.\cite{TMGesing2009}  With such a high mismatch only thin commensurate layers of BaSnO$_3$ can be grown; at a thickness of 32 nm the reported BaSnO$_3$ films are almost fully relaxed and contain high densities of TDs.\cite{SRaghavana2016}

To reduce the dislocation density in the La:BaSnO$_3$ film, we explored the insertion of an undoped BaSnO$_3$ buffer layer at the interface between the substrate (TbScO$_3$) and the La:BaSnO$_3$ film.   As the thickness of the BaSnO$_3$ layer increases, the TD density decreases as the threading component of dislocations annihilate each other, leaving behind a network of misfit dislocations.  This method is known and has already led to the highest mobilities in BaSnO$_3$  films to date; but even for thick buffer layers (330 nm), the remaining TD density is still $1.2\times10^{11}$ cm$^{-2}$.~\cite{HPaik2017}  Conceptually, growing this undoped BaSnO$_3$ buffer layer at higher substrate temperatures should lower TD densities further, but due to the significant volatility of tin oxide at substrate temperatures above about 850\degree C, this is not a viable option.
\begin{figure}[!t]
	\centering
	
		\includegraphics[width=0.47\textwidth]{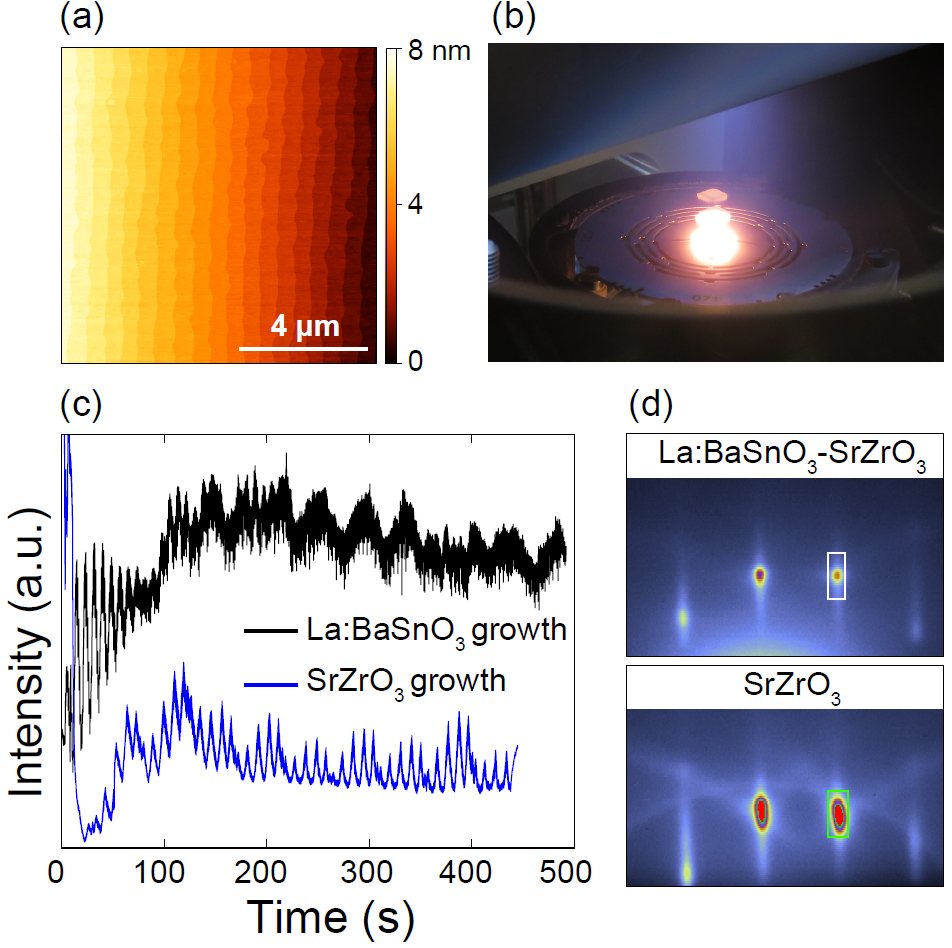}
			\caption{\linespread{1.} (a) AFM image of a typical (110) TbScO$_3$ substrate annealed \textit{in situ} at $1300^{~\circ}$C for 200 sec. (b) Inside view image of the PLD chamber showing a TbScO$_3$ substrate at $1600^{~\circ}$C with a plume ejected from the SrZrO$_3$ target during PLD growth.   Due to multiple reflections, several images  of the hot substrate are superimposed. (c) Reflection high-energy electron diffraction (RHEED) intensity oscillations during the growth of a  SrZrO$_3$ buffer layer at $1300^{~\circ}$C (blue) and La:BaSnO$_3$ film at $850^{~\circ}$C (black). (d) RHEED patterns viewed along the [110] azimuth of (top) a 25 nm thick La:BaSnO$_3$ film grown on top of (bottom) a $100$ nm thick SrZrO$_3$ buffer layer. The green and white rectangles mark the region from which  the integrated intensity as a function of time during deposition in (c) were recorded.}
		\label{fig:first}
	\end{figure}
	
In this work, we report on an alternative approach to significantly reduce the dislocation density and increase the mobility of La:BaSnO$_3$ films. By inserting between the film and the substrate an insulating buffer layer of SrZrO$_3$ grown at very high temperature, we greatly reduce the density of TDs. To avoid possible contamination, we first optimized the growth conditions of the films on (110) TbScO$_3$ substrates thermally prepared \textit{in situ} by directly heating the substrate with a CO$_2$ laser. Without resorting to post-growth treatments, we achieve a significant reduction of the density of TDs from $5\times 10^{11}$ cm$^{-2}$ to $4.9\times 10^{9} $ cm$^{-2}$ and a maximum RT mobility of 140 cm$^2$ V$^{-1}\text{s}^{-1}$ for La:BaSnO$_3$ films prepared on a SrZrO$_3$ buffer (grown at $1300^{~\circ}$C). Extending prior reports, this RT mobility is obtained for films of small thickness (25 nm). It is the highest mobility for La:BaSnO$_3$  films grown by PLD.  

Epitaxial La:BaSnO$_3$ films with a La-doping content of $6\%$ and a thickness of $25$ nm were grown on several (110) oriented TbScO$_3$ single crystalline substrates ($5\times5\times1$ mm$^3$). All samples were grown by PLD ($\lambda=248$ nm) at a target-substrate distance of $56$ mm, using a CO$_{2}$ laser substrate heating system. The La:BaSnO$_3$ films were grown at $850^{~\circ}$C and  $1.5$ J cm$^{-2}$ at $1$ Hz in $1\times10^{-1}$ mbar of O$_2$. Films were grown on either SrZrO$_3$ or BaSnO$_3$ buffer layers. The buffer layers were deposited at $4$ Hz to a thickness of $100$ nm. We chose SrZrO$_3$ as a candidate because its lattice parameter value ($4.101$ {\AA})\cite{EMete2003,AJSmith1960} is between that of La:BaSnO$_3$ ($4.116$ {\AA}) and TbScO$_3$ ($3.955$ {\AA}).\cite{WMa2018} Also, like TbScO$_3$, SrZrO$_3$ has a low vapor pressure and  can therefore be grown at high temperatures. SrZrO$_3$ layers were deposited at temperatures ranging from $850^{~\circ}$C to $1600^{~\circ}$C with a laser fluence of $2$ J cm$^{-2}$ at $1.4\times10^{-2}$ mbar of O$_2$. The BaSnO$_3$ buffer layers were grown at  $850^{~\circ}$C with a fluence of $1.5$ J cm$^{-2}$ at $1\times10^{-1}$ mbar of O$_2$. Following Ref.~\cite{HPaik2017}, BaSnO$_3$ was chosen as it has roughly the same lattice parameter as La:BaSnO$_3$. 

Prior to deposition, the TbScO$_3$  substrates were \textit{in situ}  terminated at high temperature with the CO$_2$ laser. Figure~\ref{fig:first}\textcolor{blue}{(a)} shows a typical atomic force microscopy (AFM) image of thermally prepared substrates heated at $1300^{~\circ}$C for $200$ s. The substrate surface is smooth and well-ordered, with clear and uniform terraces. More details on the \textit{in situ} thermal preparation of the substrates and related oxide surfaces are provided in Ref.~\onlinecite{MJager2019}. Figure~\ref{fig:first}\textcolor{blue}{(b)} shows a photograph of the growth chamber during the deposition of the SrZrO$_3$ buffer layer at $1600^{~\circ}$C.

The deposition of both the active layer (La:BaSnO$_3$) and buffer layer (SrZrO$_3$) were \textit{in situ} monitored by reflection high-energy electron diffraction (RHEED) [Fig.~\ref{fig:first}\textcolor{blue}{(c)}]. The SrZrO$_3$ layer was deposited at $1300^{~\circ}$C. Immediately after the deposition, the sample was cooled to  $850^{~\circ}$C at $2$ K$/$sec for the subsequent growth of the La:BaSnO$_3$ film. We observe that the intensity of the RHEED oscillations remains the same throughout the deposition of the SrZrO$_3$ layer. For the La:BaSnO$_3$ film, the intensity drops after several monolayers and then stabilizes. The intensity drop suggests a relaxation in the film after a critical thickness has been reached (see Fig. S1 of the supplementary material). The RHEED data indicate that the SrZrO$_3$ buffer layer and the La:BaSnO$_3$ films are grown in a layer-by-layer mode with a smooth surface as also demonstrated by streaky RHEED patterns for both layers [see Fig.~\ref{fig:first}\textcolor{blue}{(d)})].
\begin{figure*}[!t]
	\centering
	\includegraphics[width=1\textwidth]{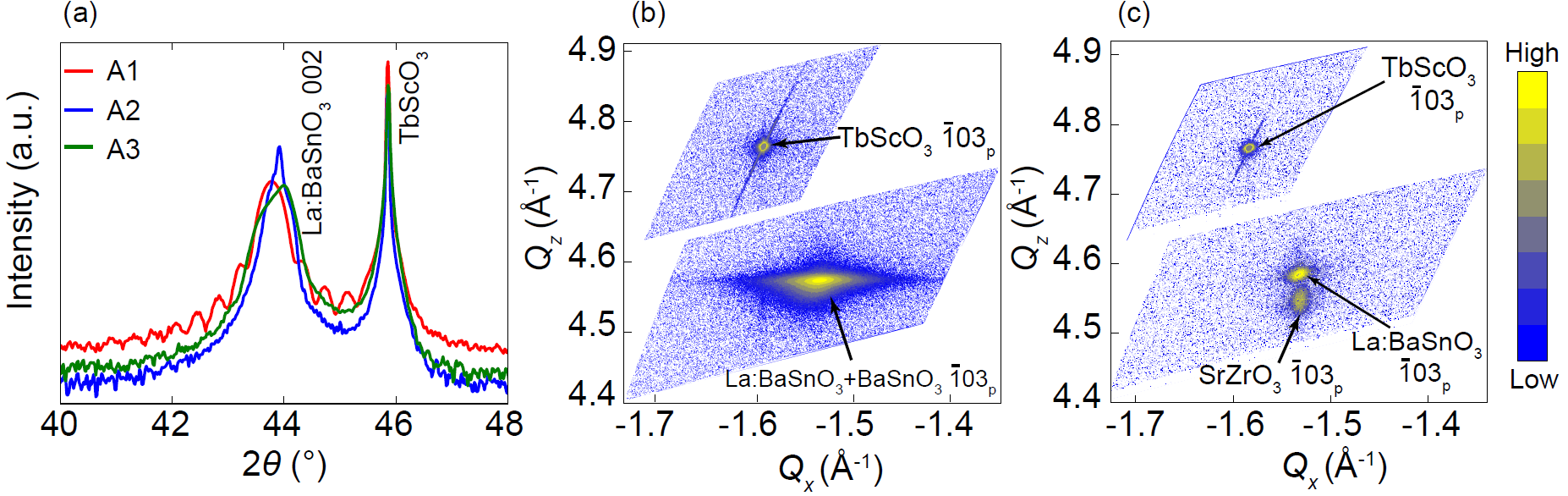}
	\caption{\linespread{1.} XRD scans of 25 nm thick La:BaSnO$_3$ films. (a) A close-up view of the $2\theta-\omega$ scan around the $002$  peak for  La:BaSnO$_3$ film grown directly on (110) TbScO$_3$ (red curve), and La:BaSnO$_3$ films deposited on top of BaSnO$_3$ (blue curve) and SrZrO$_3$ (green curve)  buffer layers grown on (110) TbScO$_3$. The $002$ peaks of  La:BaSnO$_3$ show thickness fringes. Reciprocal space maps of samples A2 and A3 in (a) around the $\bar{1}03_{\text{p}}$ reflection peaks of the (b) La:BaSnO$_3/$BaSnO$_3$ and (c) La:BaSnO$_3/$SrZrO$_3$ heterostructures, and the $\bar{1}03_{\text{p}}$ reflection of TbScO$_3$ substrates, where p refers to pseudo-cubic indices.}
	\label{fig:second}
\end{figure*}

The crystalline quality and phase purity of the films were characterized by X-ray diffraction (XRD) using Cu K$_{\alpha}$ radiation. Figure~\ref{fig:second}\textcolor{blue}{(a)} shows the $2\theta-\omega$ scans around the $002$ diffraction planes for the  La:BaSnO$_3/$TbScO$_3$ (sample A1), La:BaSnO$_3/$BaSnO$_3/$TbScO$_3$ (sample A2) and  La:BaSnO$_3/$SrZrO$_3/$TbScO$_3$ (sample A3) films. Laue thickness fringes and  phase-pure La:BaSnO$_3$  $00l$ peaks are observed, indicating smooth growth and high crystallinity [Fig.~\ref{fig:second}\textcolor{blue}{(a)}, see also Fig. S2 of the supplementary material]. The film thicknesses were extracted from the Laue thickness fringes and the Kiessig fringes observed by XRD. The extracted out-of-plane lattice parameters are $\textit{c}=4.134$ {\AA}, $4.120$ {\AA} and $4.113$ {\AA} for samples A1, A2 and A3. These values are consistent with the out-of-plane lattice constants reported previously for La:BaSnO$_3$ films\cite{WJLee2016,HMun2013} and are close to the bulk lattice parameter ($\sim4.126$ {\AA}) of polycrystalline $6\%$ La:BaSnO$_3$.\cite{THuang1995,CANiedermeier2016} Figures~\ref{fig:second}\textcolor{blue}{(b)} and \ref{fig:second}\textcolor{blue}{(c)} show reciprocal space maps (RSM) around the asymmetric ($\bar{1}03$)$_\text{p}$ reflection peaks of the films (samples A2 and A3) and the substrate. The in-plane and out-of-plane lattice constants of the film in sample A2 are $\textit{a}=4.117$ {\AA} and  $\textit{c}=4.120$ {\AA}, respectively, indicating that the film is almost completely relaxed. The relatively small deviation of the lattice constants is attributed to the large lattice mismatch ($3.9\%$) between BaSnO$_3$ and the TbScO$_3$ [Fig.~\ref{fig:second}\textcolor{blue}{(b)}]. The broadening of the film peak indicates the presence of a large density of dislocations. For sample A3, the extracted in-plane and out-of-plane lattice constants are $\textit{a}=4.103$ {\AA}, $\textit{c}=4.113$ {\AA}, indicating that the film is fully strained in the \textit{a} direction ($-0.56\%$) and partially relaxed in the \textit{c} direction. From the XRD data (see Fig. S3 of the supplementary material) of a control sample of SrZrO$_3$ grown at $1300^{~\circ}$C (in the same growth conditions as the buffer layer in sample A3), in-plane and out-of-plane lattice constants of $\textit{a}=4.085$ {\AA} and $\textit{c}=4.128$  {\AA} are obtained, indicating that the buffer layer is almost fully strained to the substrate ($3.966$  {\AA}), suggesting a small number of dislocations in the film A3 as discussed below.
	
\begin{figure}[!b]
	\centering
			\includegraphics[width=0.49\textwidth]{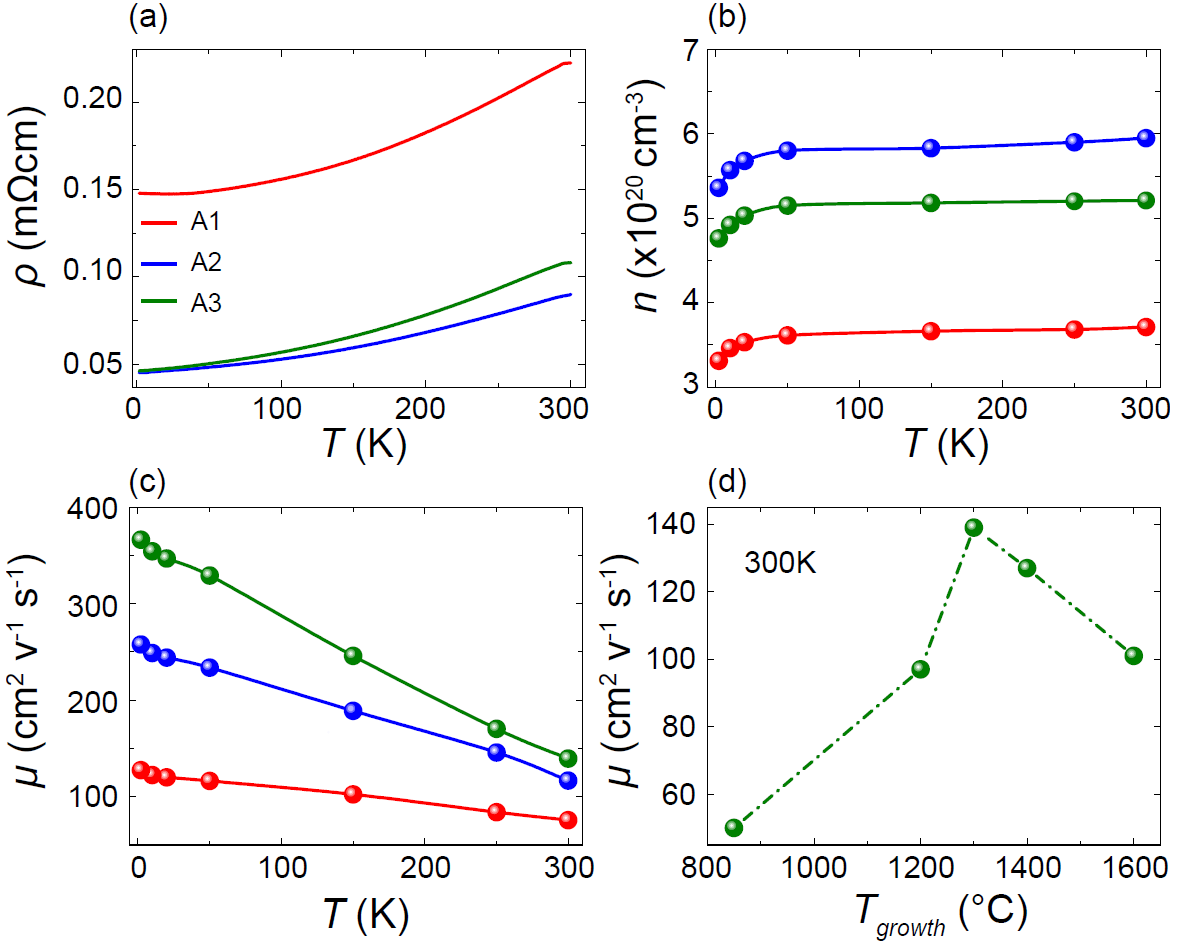}
			\caption{ Transport characteristics of 25 nm La:BaSnO$_3$ films. (a) Temperature dependence of the zero-field resistivity of the samples (red) A1, (blue) A2 and (green) A3. (b) Mobile electron carrier concentration vs. temperature and (c) electron mobility vs. temperature of the same La:BaSnO$_3$ films characterized in Figs.~\ref{fig:first}\textcolor{blue}{(c)}-\textcolor{blue}{(d)} and Fig.~\ref{fig:second}. (d) Measured electron mobility as function of the growth temperature of the SrZrO$_3$ buffer layer.}
		\label{fig:third}
	\end{figure}

Resistivity ($\rho$), electron mobility ($\mu_e$) and carrier concentration (\textit{n}) were characterized in a physical property measurement system (PPMS) in a van der Pauw geometry by wire bonding aluminum wires to the samples' corners. An excitation current of $1~\mu$A was used. Using the procedure discussed in Refs.~\cite{PNgabonziza2016,PNgabonziza2018}, the carrier concentration was calculated  as $\textit{n}=1/e\text{R}_{H}$, where $\text{R}_{H}$ is the measured Hall coefficient; and the electron mobility was extracted from $\mu_e=\text{R}_{H}/\rho$, where $\rho$ was determined by the Van der Pauw method. Figure~\ref{fig:third}\textcolor{blue}{(a)} presents the temperature dependence of the  resistivity at zero magnetic field for the samples A1, A2 and A3. All films exhibit metallic behavior over the entire temperature range, consistent with previous reports on La:BaSnO$_3$.\cite{HJKim2012-1,HJKim2012-2} 
 The temperature dependence of the carrier density shows the bahavior of a degenerate semiconductor.~\cite{HJKim2012-1} The concentration of the negatively charged carriers  is temperature independent at high temperatures, but starts to decrease below 50 K following the freeze out at the La$^{+3}$ ions [Fig.~\ref{fig:third}\textcolor{blue}{(b)}].~\cite{PVWadekar2014}  The temperature dependence of the electron mobility is presented in Fig.~\ref{fig:third}\textcolor{blue}{(c)}. For all three samples, $\mu_e$ increases down to the lowest temperature ($2$ K), contrary to previous studies which reported a  saturation below 50 K.\cite{SRaghavana2016,HPaik2017} This behavior suggests a significant reduction of phonon scattering for these films at low temperatures. 
 For the sample directly grown on the substrate (A1), a mobility of 76 cm$^2$ V$^{-1}\text{s}^{-1}$ (RT) was measured. In the sample with the BaSnO$_3$ buffer layer (A2), the RT $\mu_e$ is improved to 117 cm$^2$ V$^{-1}\text{s}^{-1}$, but apparently still limited by the high density of TDs, as discussed below. To reduce the density of TDs, we switched to SrZrO$_3$ buffer layers. Several La:BaSnO$_3/$SrZrO$_3$ heterostructures were prepared by varying the growth temperatures of the SrZrO$_3$ layer from $850^{~\circ}$C to $1600^{~\circ}$C, while keeping the other growth parameters constant. We find $1300^{~\circ}$C as the optimal growth temperature of the SrZrO$_3$ buffer layer to achieve high mobility in the La:BaSnO$_3$ layers [Fig.~\ref{fig:third}\textcolor{blue}{(d)}].   Above $1300^{~\circ}$C the buffer layers start to become more non-stoichiometric, i.e., the ratio of Sr:Zr  deviating from 1 in SrZrO$_3$ due to the volatility of constituents~\cite{Engel-Herbert2013} (see Fig. S3(d) in supplemental materials). This results in an enhanced defect density (e.g. shear defects and/or point defects arising from nonstoichiometry~\cite{HPaik2017}), limiting the mobility of the La:BaSnO$_3$ layers deposited on buffer layers grown at temperatures above $1300^{~\circ}$C.	The sample A3 prepared on a buffer layer grown at $1300^{~\circ}$C has the highest mobility $\mu_e = 140$ cm$^2$ V$^{-1}\text{s}^{-1}$ (RT) and $375$ cm$^2$ V$^{-1}\text{s}^{-1}$ ($2$ K) [Fig.~\ref{fig:third}\textcolor{blue}{(c)}]. For PLD grown La:BaSnO$_3$ thin films, this RT $\mu_e$ is $\sim30\%$ higher than the previous record ($100$ cm$^2$ V$^{-1}\text{s}^{-1}$) achieved without post-growth treatment on BaSnO$_3$ substrates;\cite{WJLee2016} and $15\%$ higher than reported enhanced mobility ($122$ cm$^2$ V$^{-1}\text{s}^{-1}$) after post-annealing processes in H$_2$ forming gas at $950^{~\circ}$C on SrTiO$_3$.\cite{DYoon2018} 
	
\begin{figure}[!t]
	\centering
	
		\includegraphics[width=0.49\textwidth]{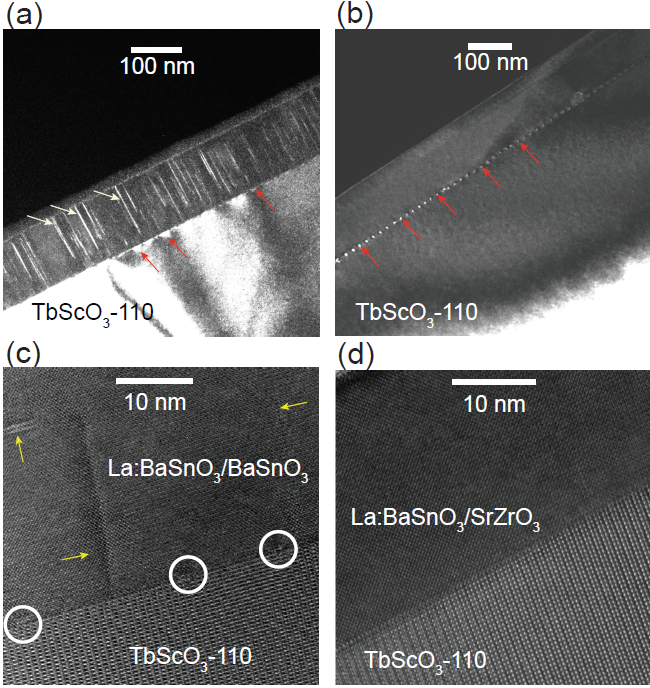}\caption{Weak-beam dark-field TEM images of samples (a) A2 and (b) A3. Misfit dislocations along the interfaces are shown by red arrows. Edge-type threading dislocations are visible in (a), indicated by vertical bright contrasts (white arrows). Only periodic misfit dislocations are visible in (b). Their average distance is $17$ nm. High-resolution TEM images for the same  heterostructures (c) A2 and (d) A3. Misfit dislocations indicated by white circles in (c) suggest a fully relaxed interface. Stacking faults (yellow arrows) are also visible. An almost fully strained interface is seen in (d) with no apparent structural defects.}
		\label{fig:fourth}
	\end{figure}

To further investigate the role of TDs that act as scattering centers and trap electrons,\cite{APrakash2017,HMun2013,JShin2016,SYu2016} we studied the defect structure of the samples A2 and A3 using cross-sectional transmission electron microscopy (TEM). Figures~\ref{fig:fourth}\textcolor{blue}{(a)} and \ref{fig:fourth}\textcolor{blue}{(b)} present the weak-beam dark-field TEM (WB-DFTEM) images of the entire film thickness for the samples A2 and A3, respectively. For both samples misfit dislocations are visible along the interface as bright dots, indicated in figures~\ref{fig:fourth}\textcolor{blue}{(a)} and \ref{fig:fourth}\textcolor{blue}{(b)}  by red arrows. To accommodate the large lattice mismatch between the substrate and the film, these misfit dislocations generate edge-type defects, which extend vertically through the film (bright contrast indicated by white arrows in Fig.~\ref{fig:fourth}\textcolor{blue}{(a)}).  For a TEM specimen of $16$ nm thick, the extracted density of TDs in sample A2 was $5\times 10^{11}$ cm$^{-2}$. This density of TDs is in agreement with previous reports on La:BaSnO$_3$ films prepared on DyScO$_3$\cite{HPaik2017} and SrTiO$_3$\cite{HMun2013} substrates. Interestingly, TDs were barely observed in the highest mobility sample A3 as shown in the WB-DFTEM image in Fig.~\ref{fig:fourth}\textcolor{blue}{(b)}. Even for a TEM specimen of $205.5$ nm thick, the TDs density was $4.9\times 10^{9}$ cm$^{-2}$ which is two orders of magnitude lower than the one in the sample A2 and previously reported TD densities for La:BaSnO$_3$ films.\cite{HPaik2017,HMun2013} Figures~\ref{fig:fourth}\textcolor{blue}{(c)} and \ref{fig:fourth}\textcolor{blue}{(d)} depict the high resolution TEM images of samples A2 and A3, respectively. A fully relaxed interface between the TbScO$_3$ substrate and the BaSnO$_3$ layer is seen in sample A2 as indicated by white circles. Additional structural defects such as stacking faults are also visible in the film [see yellow arrows in Fig.~\ref{fig:fourth}\textcolor{blue}{(c)}]. On the other hand, a strained interface is seen in sample A3 with no apparent structural defects [Fig.~\ref{fig:fourth}\textcolor{blue}{(d)}].
	
The low density of TDs in sample A3 is attributed to the high temperature used for the SrZrO$_3$ buffer layer growth. One of the effective ways of minimizing the TDs density is to enhance the motion and reaction of TDs by high thermal stress. This method was demonstrated to be efficient in epitaxial semiconductor GaAs films on Si substrates.\cite{YTakagi1994,DDeppe1988} At high temperatures, the velocity of the dislocations glide motion and the concentration of vacancies that support the climb motion of the dislocations are exponentially enhanced.\cite{IYonenaga1987} Thus, the observed significant reduction of the TDs density in sample A3 is attributed to the enhancement of the glide and climb motion at high temperatures. Stress that increases the glide motion originates from the difference in the thermal expansion coefficients of $\sim3.1$ between SrZrO$_3$ and TbScO$_3$\cite{RUecker2008,DdeLigny1996} and also possible grown-in strain in the SrZrO$_3$ layer. The latter is supported by the fact that SrZrO$_3$ undergoes a phase transition from tetragonal to cubic at high temperature,\cite{DdeLigny1996} indicating epitaxial strain at the interface during cool down, yielding a reduction of the lattice mismatch. Our results suggest that the high temperature grown SrZrO$_3$ epilayer deposited not only on TbScO$_3$ but also on other oxide substrates (e.g. SrTiO$_3$, DyScO$_3$, MgO) can be utilized as a template for subsequent growth of high mobility La:BaSnO$_3$ films with fewer TDs. 

Although we have demonstrated that by inserting a high-temperature grown SrZrO$_3$  buffer layer between the film and the substrate the density of TDs significantly reduces with an associated improvement of the electron mobility in La:BaSnO$_3$ films, we point to several observations. Normally, scattering by TDs not only diminishes the electron mobility, but also reduces the number of free charge carriers~\cite{HMun2013}. However, we observe that the carrier density of sample A2 (with the high density of TDs) is higher than sample A3 (with a low TD density) [Fig.~\ref{fig:third}\textcolor{blue}{(b)}]. Given that the active layer in our samples is thin (25 nm), the lower number of free charge carriers in sample A3 suggests that not only TDs are trapping electrons, but that also  effects such as surface scattering or interface traps are lowering the density of mobile carriers. For the sample A2, these contributions are expected to be less pronounced as the buffer layer (BaSnO$_3$) and the active layer consist of the same materials.

In summary, we explore the possible use of buffer layers grown at very high temperatures for the reduction of TDs and improvement of electron mobility in epitaxial La-doped BaSnO$_3/$SrZrO$_3$ heterostructures grown on (110) TbScO$_3$ substrates. For La:BaSnO$_3$ films prepared on a SrZrO$_3$ buffer layer grown at $1300^{~\circ}$C a RT mobility of 140 cm$^2$ V$^{-1}\text{s}^{-1}$ has been achieved, together with a reduction of the density of TDs to $4.9\times 10^{9}$ cm$^{-2}$, all without post-growth sample treatment. 
With the insertion of a high temperature grown buffer layer between the La:BaSnO$_3$ film and the TbScO$_3$ substrate, the mobility has been doubled, which opens a new road towards high mobilities in La:BaSnO$_3$ based electronic devices. \\

We acknowledge valuable discussions with Darrell G. Schlom and Hans Boschker, and thank Helga Hoier and Marion Hagel for technical assistance.

\clearpage
\onecolumngrid

\setcounter{table}{0}
\renewcommand{\thetable}{S\arabic{table}}%
\setcounter{figure}{0}
\renewcommand{\thefigure}{S\arabic{figure}}%
\setcounter{equation}{0}
\renewcommand{\theequation}{S\arabic{equation}}%

\begin{center}
\title*{\textbf{\Large{Supplementary information:\newline \newline High-temperature-grown buffer layer boosts electron mobility in epitaxial La-doped BaSnO$_3$/SrZrO$_3$ heterostructures}}}
\end{center}
\begin{center}
\large{Arnaud  P. Nono Tchiomo$^{1,2}$, Wolfgang Braun$^{1}$, Bryan P. Doyle$^{2}$, Wilfried Sigle$^{1}$, Peter van Aken$^{1}$, Jochen Mannhart$^{1}$, and Prosper Ngabonziza$^{1,2}$\newline \newline\newline
$^{1)}$\textit{Max Planck Institute for Solid State Research, Heisenbergstr. 1, 70569} \\ \textit{Stuttgart, Germany}\\
$^{2)}$\textit{Department of Physics, University of Johannesburg, P.O. Box 524 Auckland Park 2006,} \\ \textit{Johannesburg, South Africa}}
\end{center}
\begin{center}
\Large{\textbf{Analysis of RHEED intensity oscillations}}
\end{center}
Reflection high-energy electron diffraction (RHEED) data were acquired using the Safire data acquisition software. The RHEED gun was a differentially pumped Staib system operated at 30 keV. To fit the RHEED raw data from our samples, we followed the procedure for the analysis of the RHEED intensity oscillations presented in Refs.~\cite{Braun1,Braun2}. We show below RHEED data of the La:BaSnO$_3$ active layer and SrZrO$_3$ buffer layer for the highest mobility sample A3 discussed in the main text.\\
\begin{figure*}[!h]
\begin{center}
{\includegraphics[width=0.75\textwidth]{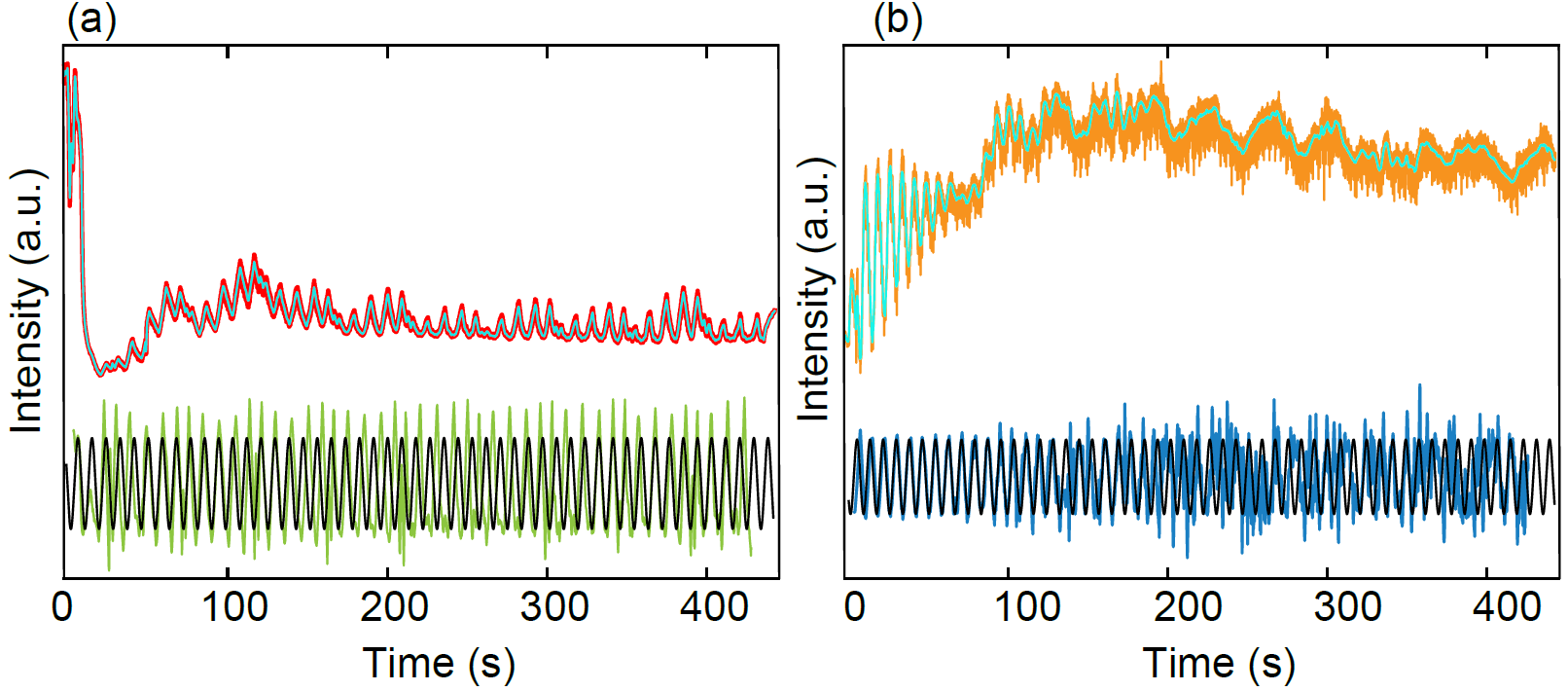}}\end{center}      
\begin{flushleft}\textbf{Figure S1:} RHEED intensity oscillations during the growth of the sample A3 discussed in the main text. (a) Intensity oscillations during the growth of SrZrO$_3$ buffer layer (red curve) together with a best fit to the raw data (cyan curve). The bottom green curve was obtained by using an algorithm for RHEED data analysis from Ref.~[\textcolor{blue}{1}] and the transformed data were then fitted by a sine function (black curve). (b) The same analysis was performed on the 25-nm-thick La:BaSnO$_3$ grown on top of the SrZrO$_3$ buffer layer.
\end{flushleft}
\end{figure*}

\begin{center}
\Large{\textbf{X-ray diffraction measurements}}
\end{center}
We present here additional x-ray diffraction (XRD) measurements over a wider region for the samples A1 (La:BaSnO$_3$/TbScO$_3$), A2 (La:BaSnO$_3$/BaSnO$_3$/TbScO$_3$) with the buffer layer grown at 850$^{~\circ}$C and A3  (La:BaSnO$_3$/SrZrO$_3$/TbScO$_3$) with the buffer layer grown at 1300$^{~\circ}$C. Data from these samples are discussed in the main text. Only phase-pure 00\textit{l} peaks from the film are observed, underlining the crystalline quality of the samples.
	\begin{figure*}[!h]
	\begin{center}
	{\includegraphics[width=0.8\textwidth]{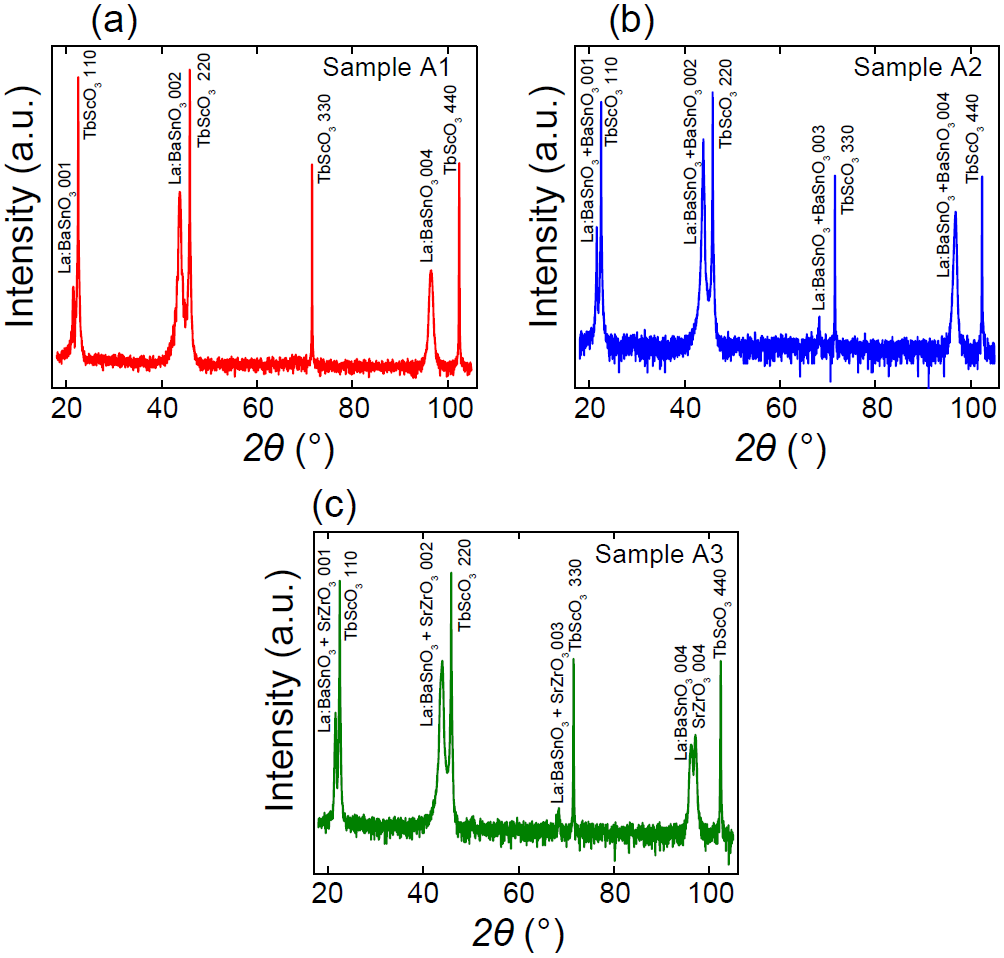}}   
	\end{center}
\begin{flushleft}
   \textbf{Figure S2:} XRD measurements. $2\theta-\omega$ scans over a wider region for the samples (a) A1, (b) A2 and (c) A3 showing only phase-pure 00\textit{l} peaks from the samples.
	 \end{flushleft}	      
\end{figure*}
\\ \\
\begin{center}
\Large{\textbf{Characterization of SrZrO$_3$ buffer layer}}
\end{center}
XRD and atomic force microscopy (AFM) data for a control sample of SrZrO$_3$ grown at 1300$^{~\circ}$C are presented. The sample was prepared in the same growth conditions as the buffer layer in the sample A3 discussed in the main text. The lattice mismatch between SrZrO$_3$ and TbScO$_3$ from this control sample grown at $1300^{~\circ}$C was reduced from $\sim 3.7\%$ to $\sim 3\%$.  As suggested by the SrO partial pressure vs. temperature diagram (see Fig.~\textcolor{blue}{S3 (d)}),
the stoichiometry of the SrZrO$_3$ buffer layer deteriorates at very high growth temperatures (above $1300^{~\circ}$C) due to the volatility of SrO.
\begin{figure}[!h]
\centering
{\includegraphics[width=0.69\textwidth]{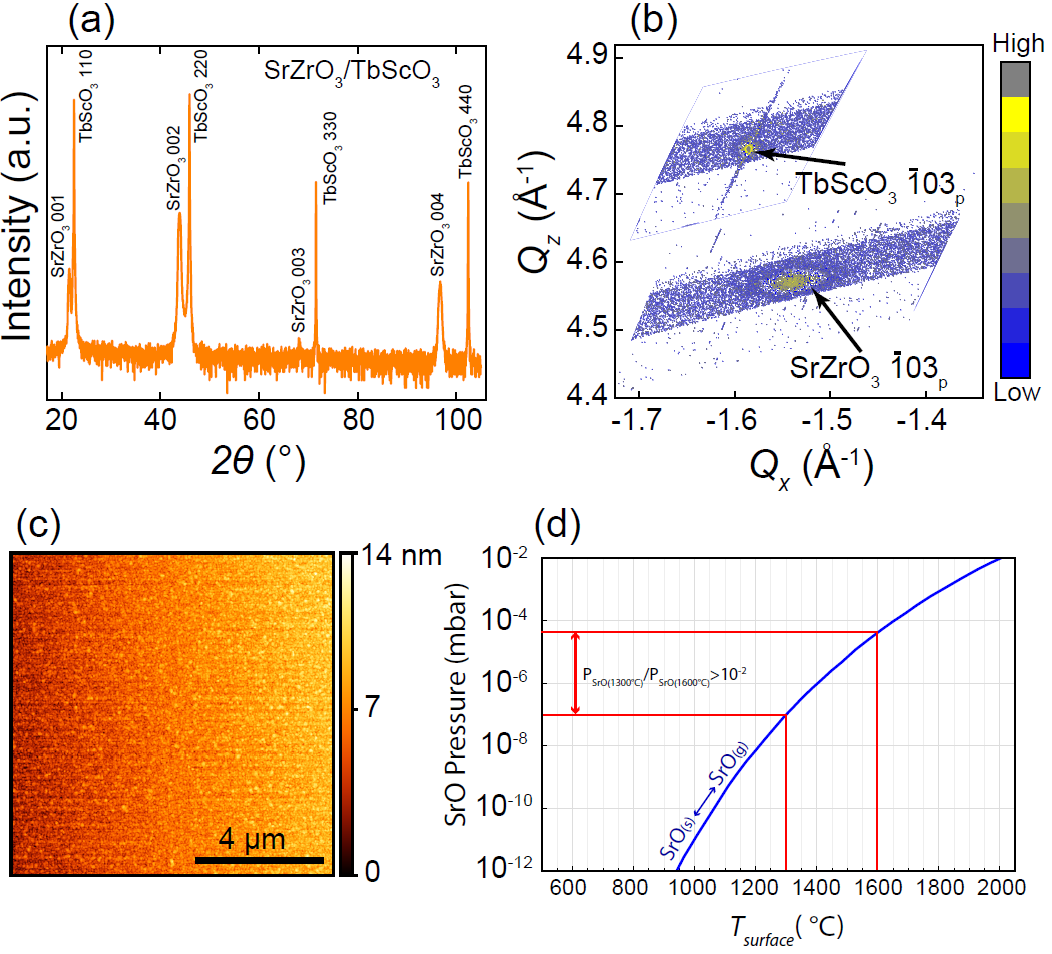}}
\begin{flushleft}
\textbf{Figure S3:} Characterization of the SrZrO$_3$ buffer layer. (a) $2\theta-\omega$ scan over a wider region for a 100 nm thick SrZrO$_3$ buffer layer grown at 1300$^{~\circ}$C on a TbScO$_3$ substrate. Only the phase-pure 00\textit{l} peaks from the SrZrO$_3$ film are observed. (b) Reciprocal space map and (c) AFM image of the same SrZrO$_3$ film. (d) SrO phase diagram (adapted from Ref.~\cite{Engel-Herbert2013}).
\end{flushleft}
	\label{fig:S3}
\end{figure}

\begin{center}
\end{center}
\begin{table*}[!h]
\begin{center}
	\caption{Overview of the extracted transport characteristics (mobility and carrier density) and lattice parameters for different La:BaSnO$_3$ ($25$ nm)/SrZrO$_3$ ($100$ nm)/TbScO$_3$ heterostructures for several growth temperatures of the SrZrO$_3$ buffer layer.}
	\begin{tabular}{||c|c|c|c|c||}
		\hline\hline
		\scriptsize{SrZrO$_3$ growth} & & &  & \\
		\scriptsize{temperature} & \scriptsize{In-plane lattice constant} & \scriptsize{Out-of-plane lattice constant } & \scriptsize{Carrier density at 300 K}  & \scriptsize{Mobility at 300 K}\\
		& \scriptsize{La:BaSnO$_3$} &\scriptsize{La:BaSnO$_3$} &  & \\
		\scriptsize{($^{\circ}$C)} & \scriptsize{\textit{a} ({\AA})} &\scriptsize{\textit{c} ({\AA})} & \scriptsize{(cm$^{-3}$)} & \scriptsize{(cm$^2$ V$^{-1}$ s$^{-1}$)} \\
		\hline
		\scriptsize{850} & \scriptsize{4.09} & \scriptsize{4.136} & \scriptsize{$1.8\times 10^{20}$} & \scriptsize{50} \\
		\hline
		\scriptsize{1200} & \scriptsize{4.07} & \scriptsize{4.14} & \scriptsize{$4.6\times 10^{20}$} & \scriptsize{97.4} \\
		\hline
		\scriptsize{1300} & \scriptsize{4.103} & \scriptsize{4.113} & \scriptsize{$5.2\times 10^{20}$} & \scriptsize{140} \\
		\hline
			\scriptsize{1300} & \scriptsize{4.103} & \scriptsize{4.113} & \scriptsize{$5.0\times 10^{20}$} & \scriptsize{136.2} \\
		\hline
		\scriptsize{1400} & \scriptsize{4.10} & \scriptsize{4.115} & \scriptsize{$4.7\times 10^{20}$} & \scriptsize{127} \\
		\hline
		\scriptsize{1600} & \scriptsize{4.06} & \scriptsize{4.142} & \scriptsize{$5\times 10^{20}$} & \scriptsize{100.8} \\
		\hline\hline
	\end{tabular}
\end{center} 
\end{table*} 

\end{document}